\documentclass[10pt,letterpaper,compsoc,conference]{iiswc26}

\usepackage{cite}
\usepackage{amsmath,amssymb,amsfonts}
\usepackage{algorithmic}
\usepackage{graphicx}
\usepackage[dvipsnames]{xcolor}
\usepackage[final]{microtype}
\usepackage[italic]{mathastext}
\usepackage{libertine}
\usepackage[T1]{fontenc}
\usepackage{textcomp}
\usepackage[varqu,varl]{zi4}
\usepackage[all]{nowidow}
\usepackage[auth-lg,affil-it]{authblk}
\usepackage{flushend}
\usepackage{fancyhdr}

\usepackage{url}
\usepackage{makecell}
\usepackage{kotex} 
\usepackage{booktabs}
\usepackage{multirow}
\usepackage{tabularx}
\usepackage{ragged2e}  
\newcolumntype{Y}{>{\RaggedRight\arraybackslash}X}
\usepackage{dblfloatfix}

\usepackage{caption}
\usepackage{subcaption}
\providecommand{\cmark}{\ensuremath{\checkmark}}
\providecommand{\xmark}{\ensuremath{\times}}
\usepackage{enumitem}

\begin{document}


\title{Latency Prediction for LLM Inference on NPU Systems}




\author[1]{Juhyun Park}
\author[1]{Seungwoo Jeong}
\author[2]{Jingyu Lee}
\author[1]{Kyungyong Lee\thanks{Corresponding author: kyungyong@hanyang.ac.kr}}
\affil[1]{Department of Data Science, Hanyang University, Seoul, Republic of Korea}
\affil[2]{Department of Artificial Intelligence, Hanyang University, Seoul, Republic of Korea}
\affil[ ]{\{anchovypark, swjeong25, ljk8540009, kyungyong\}@hanyang.ac.kr}

\maketitle
\pagestyle{plain}


\begin{abstract}
Deploying Large Language Models (LLMs) requires exploring a large configuration space spanning parallelization strategies, batching techniques, and scheduling policies. Exhaustive measurement across this space is impractical, making latency prediction essential for system optimization. While NPUs have emerged as accelerators designed for LLM inference, no prediction methodology has been established for them. Specifically, applying prior work to LLM inference latency prediction on NPUs faces three challenges: undisclosed microarchitecture of commercial NPUs, unpredictable compiler optimizations, and latency non-linearity induced by bucketing.

We present LENS, a latency estimator that predicts NPU inference latency without information on the microarchitecture or compiler, and captures the non-linear latency induced by bucketing. LENS profiles each bucket with two end-to-end (E2E) measurements and composes the results to predict latency for arbitrary input-output length combinations. We validate LENS across NPUs from multiple vendors, several LLMs, and diverse workloads, achieving a mean prediction error of 2.15\%. We further compare LENS against two methodologically related baselines, confirming the validity of its approach.
\end{abstract}

\section{Introduction}
\label{sec:intro}


The Transformer architecture~\cite{transformer} has driven rapid advances in Large Language Models (LLMs), and GPUs have served as the primary accelerator for this progress. LLM inference, however, exposes a structural mismatch with GPU architecture. The decode phase reads the entire accumulated KV cache from memory at every step, making memory bandwidth the dominant bottleneck~\cite{splitwise, distserve}. In addition, matmul and non-matmul kernels execute sequentially on the same SM, limiting hardware utilization~\cite{orca, splitwise}. These limitations have led to the emergence of NPUs, domain-specific accelerators designed for LLM inference~\cite{tpu-v1}. NPUs currently serve as production infrastructure for LLM deployment, including Anthropic's Claude~\cite{anthropic-tpu, anthropic-trainium} and Google's Gemini~\cite{google-trillium}.

Operating NPU-based serving systems requires exploring a vast configuration space spanning accelerator selection, batch size, and parallelization strategies. On NPUs, this exploration is further amplified by static compilation, which requires a separate compilation for each configuration. Despite this need, no prediction methodology has been established for NPUs. 

To enable LLM inference latency prediction on NPUs, we identify three challenges that block the direct application of prior work. First, the microarchitectural details of commercial NPUs are not disclosed at the level required for cycle-accurate simulation. Simulation-based NPU studies~\cite{onnx-sim, pytorch-sim} require detailed microarchitectural information as input, including systolic array dimensions, memory hierarchy, and Network-on-Chip (NoC) topology, yet recent commercial NPUs do not sufficiently disclose such information. 

Second, the compilers of commercial NPUs apply graph optimizations in unpredictable ways. The physical separation of heterogeneous engines drives NPUs to form kernel fusion and optimization patterns that differ qualitatively from those of GPUs. When GPU-based methodologies~\cite{vidur, neusight, llmservingsim2.0, habitat}, which sum latencies measured at pre-defined units, are directly applied to NPUs, error rates of up to 493\% are observed.

Third, the static compilation of NPUs renders latency a discontinuous function of sequence length. Since each input shape requires a separate compilation, NPU-based serving systems adopt a bucketing strategy that pads inputs to predefined lengths, causing latency to jump discretely at bucket boundaries. This step-function distribution invalidates the assumption of existing prediction methodologies, which interpolate continuous functions between measurement points.

To address these challenges, we leverage two observations: the end-to-end (E2E) latency of a compiled binary is externally observable, and all inputs within a bucket share the same compiled binary. Building on these observations, we present LENS (Latency Estimator for NPU Systems), which predicts LLM inference latency for arbitrary input-output length combinations from only two E2E measurements per bucket.

\begin{itemize}
\item Characterization of three challenges that obstruct LLM inference latency prediction on commercial NPUs.
\item Reinterpretation of the bucketing structure inherent in NPU serving systems as a natural unit of measurement for prediction, together with the design of LENS, a prediction tool that treats NPUs as black boxes and operates solely on per-bucket E2E profiling.
\item Validation of LENS across NPUs from multiple vendors, several LLMs, and diverse workloads, achieving a mean prediction error of 2.15\%, and comparison against two methodologically related baselines.
\item Demonstration through two case studies that NPU configuration optima cannot be reasoned about without direct measurement, with LENS supporting this search through per-bucket E2E profiling.
\end{itemize}


\section{NPU for LLM Serving Systems}
\label{sec:background}

NPUs are hardware accelerators designed for LLM inference~\cite{tpuv4-2023, llm.npu, shadownpu}. This section examines the architectural design of NPUs for this workload, and the research direction needed for their effective use.

\subsection{LLM Inference on GPUs}
\label{sec:bg_npu_arch}

Workloads dominated by large-scale matrix multiplication, such as LLM training, are executed efficiently on the massively parallel compute cores of GPUs~\cite{megatron_LM, grattafiori2024llama3herdmodels}. Modern GPUs further accelerate this operation through dedicated Tensor Cores. This alignment has driven the success of GPU-based LLM training.


LLM inference, however, splits into two phases with distinct computational characteristics due to the Key/Value (KV) cache. Prefill processes the entire input prompt in parallel, and matrix multiplication still dominates this phase. Decode then generates output tokens one at a time autoregressively. By reusing cached KV tensors from previous iterations, each decode iteration processes solely the newly generated token, and matrix-vector multiplication, which has low arithmetic intensity, dominates this phase. 


These characteristics of LLM inference expose two structural mismatches with GPU architecture. First, the low arithmetic intensity renders decode memory-bound, and the parallel compute cores of GPUs remain underutilized~\cite{distserve, splitwise}. Second, Tensor Cores and CUDA Cores are co-located on the same SM, which forces matmul and non-matmul kernels to execute serially. With the runtime share of matrix multiplication reduced relative to prefill, non-matmul kernels (e.g., element-wise operations, activations, normalizations) account for a substantial fraction of decode latency~\cite{splitwise, sarathi-serve}. During decode, matrix-vector multiplication occupies the Tensor Cores in a memory-bound state, and the colocated CUDA Cores remain idle for the duration of the kernel. Non-matmul kernels cannot be dispatched to these idle resources and must await completion of the matmul kernel. The GPU therefore fails to reach peak throughput during inference~\cite{mind-the-memory-gap}.



\subsection{NPUs for LLM inference}
\label{NPU design}

\begin{figure}[t]
\centering
\includegraphics[width=\columnwidth]{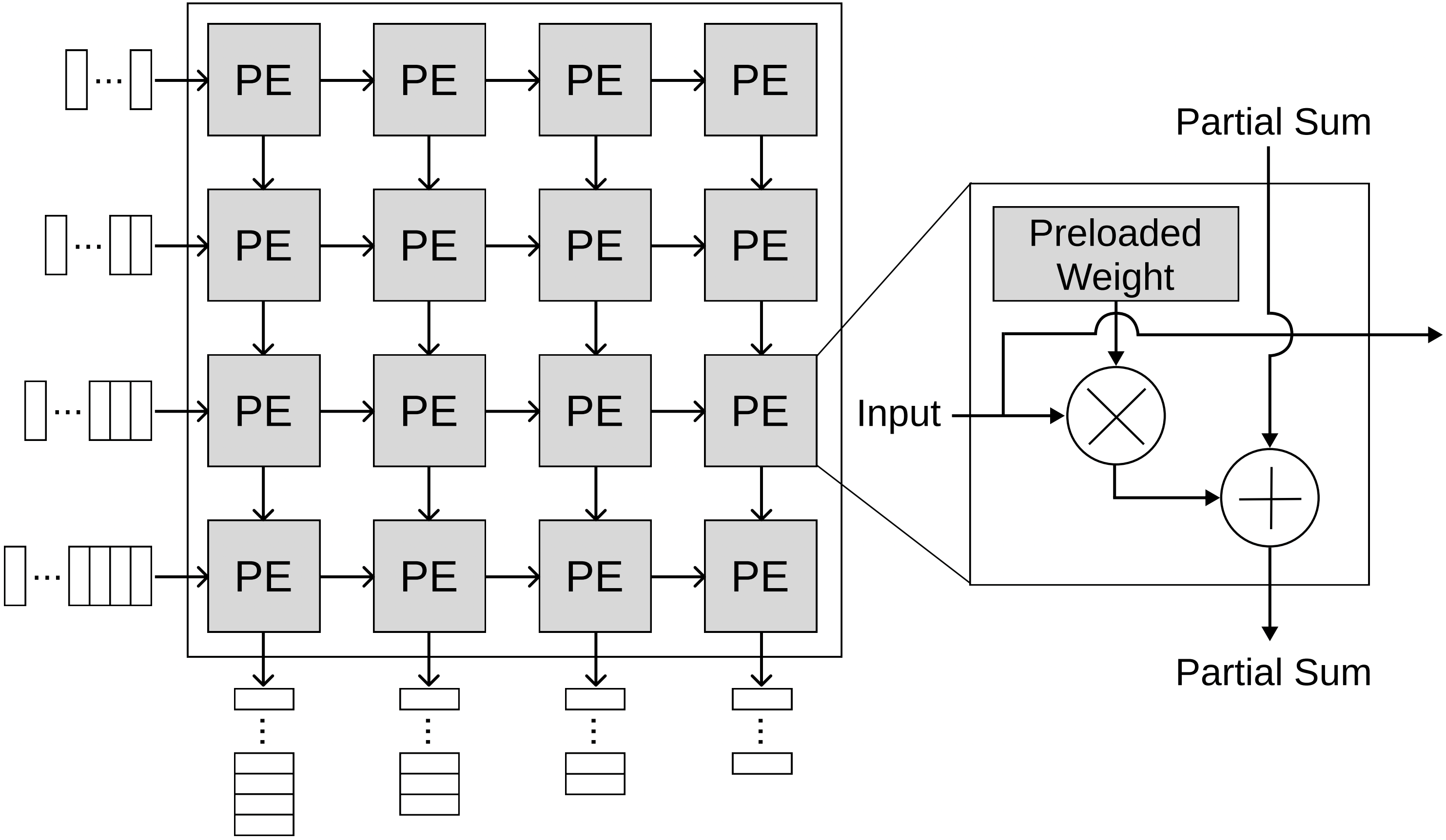}
\caption{Structure of a systolic array.}
\label{fig:npu-systolic_array}
\end{figure}

The mismatch between GPU architectures and LLM inference has motivated Neural Processing Units (NPUs)~\cite{tpu-v1}. NPUs address both mismatches — compute underutilization under low arithmetic intensity, and forced serialization of heterogeneous kernel execution — through architectural choices specialized for LLM inference~\cite{tpuv4-2023, llm.npu, shadownpu}. First, NPUs adopt dataflow architectures that maximize on-chip data reuse at the hardware level. Figure~\ref{fig:npu-systolic_array} illustrates the architecture of a systolic array under a weight-stationary dataflow. Data streams sequentially through processing elements (PEs), and the operands (i.e., weights) resident in each PE are reused across multiple computations, minimizing off-chip memory accesses. This hardware-level data reuse pattern effectively alleviates the memory bandwidth bottleneck of the memory-bound decode phase.



\begin{figure}[t]
\centering
\includegraphics[width=\columnwidth]{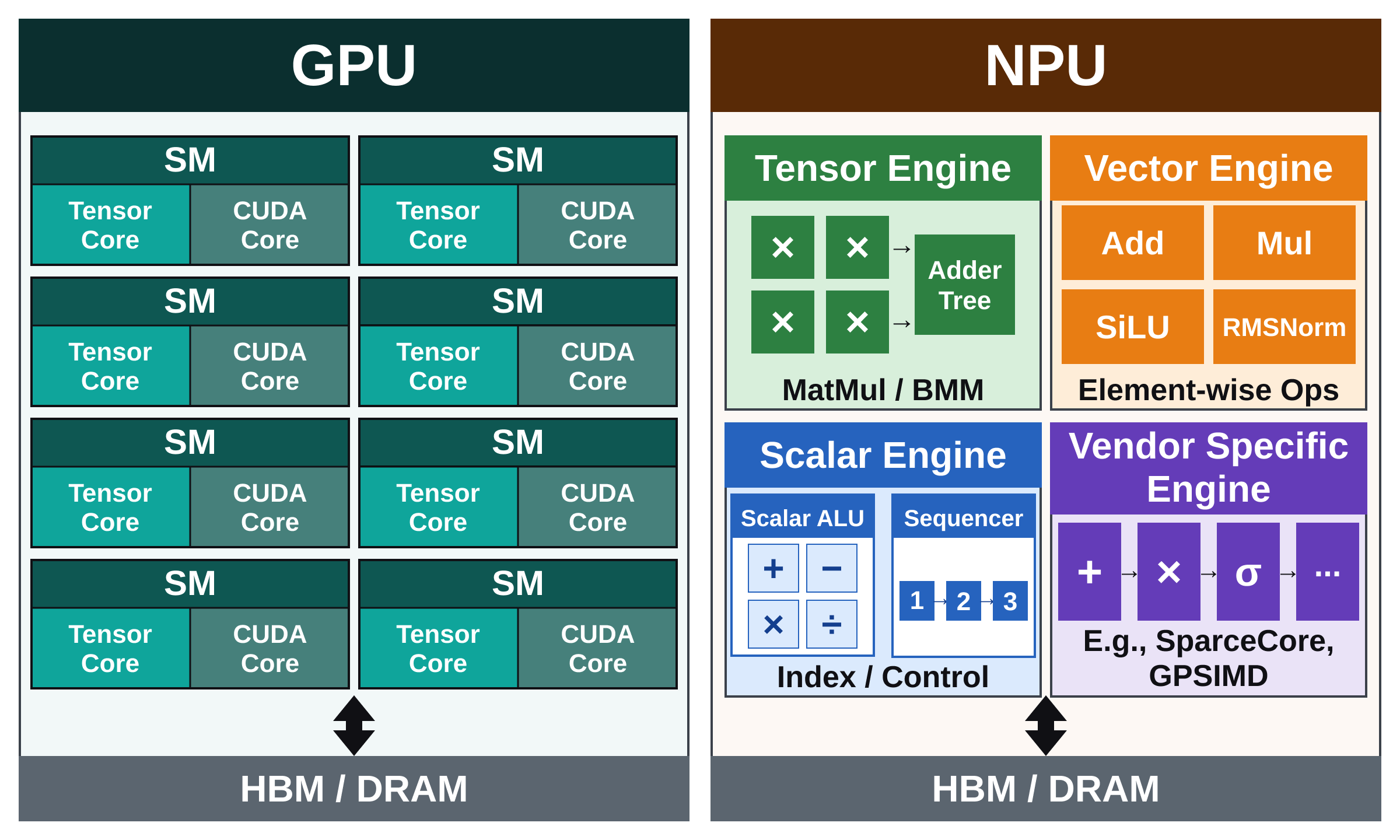}
\caption{A representative architecture of an NPU.}
\label{fig:npu-architecture}
\end{figure}

Second, GPUs execute all operations on shared SMs regardless of operation type, whereas NPUs employ physically separate engines for each operation type. Figure~\ref{fig:npu-architecture} contrasts the architectures of GPUs and NPUs. An NPU is typically partitioned into a Tensor Engine, which handles matrix multiplications; a Vector Engine, which handles element-wise operations, softmax, and normalization; and a Scalar Engine, which handles index computation and loop counters. Each NPU may further include vendor-specific engines (e.g., Google TPU's SparseCore, AWS NeuronCore's GPSIMD). These engines are physically separated, and the output of one engine can be pipelined directly into the input of another through on-chip pathways without traversing HBM. The separate-engine architecture enables heterogeneous operations to execute concurrently, thereby minimizing the device underutilization that arises from serial execution.

\subsection{LLM Inference Latency Prediction}
\label{sec:bg_latency_pred}

The deployment of LLM serving systems must satisfy a Service Level Objective (SLO) that bounds the response latency for user requests, while maximizing throughput under a limited budget. Meeting these goals involves multiple design decisions, including the selection of accelerator, batch size, and parallelization strategies such as Tensor Parallelism (TP) and Pipeline Parallelism (PP), each of which entails trade-offs among latency, throughput, and cost. Identifying the optimal configuration within this multidimensional trade-off space is therefore essential.

However, exploring this configuration space is prohibitively expensive in practice, as the number of candidate configurations explodes combinatorially across the design dimensions. Accurately characterizing the performance of each candidate configuration requires deploying and executing the model on the target platform, rendering exhaustive measurement infeasible. Vidur~\cite{vidur} reports that exploring the optimal deployment configurations of LLaMA2-70B across two GPUs (A100 and H100) and three workloads requires 42{,}000 GPU-hours, highlighting the necessity of latency prediction. 

Despite this necessity, predicting LLM inference latency is non-trivial. The decode phase incurs computation and memory access volumes that grow non-linearly with input and output lengths, since the KV cache accumulates at each step. These dynamics combine with batch size, sequence length, hardware utilization, and bandwidth contention at runtime. Analytical models that treat latency as a simple function of input variables therefore incur substantial errors. NeuSight~\cite{neusight} demonstrates that applying the linear regression-based prediction approach~\cite{Linear-Regression-DNN} from conventional DNN settings to LLM inference yields an error rate of approximately 61.2\%.

To address these challenges, various prediction methodologies have been developed for GPU environments, as summarized in Table~\ref{tab:gpu_prediction}. While these approaches differ in their specifics, they share a common formulation that decomposes LLM inference into pre-defined units and estimates E2E latency by summing the per-unit latencies. In contrast, no such prediction methodology has been established for NPUs, despite the higher exploration cost imposed by their static compilation characteristics. We address this gap with LENS, a latency prediction methodology for NPUs.


\begin{table}[t]
  \centering
  \caption[Prior work on LLM inference latency prediction in GPU environments]{%
    \textbf{Prior work on LLM inference latency 
    prediction in GPU environments.}
    Categorized by the granularity of decomposition.}
  \label{tab:gpu_prediction}
  \footnotesize
  \setlength{\tabcolsep}{2pt}
  \begin{tabularx}{\linewidth}{@{}>{\RaggedRight\arraybackslash}p{0.3\linewidth} 
                                  >{\RaggedRight\arraybackslash}p{0.3\linewidth} 
                                  >{\RaggedRight\arraybackslash}X@{}}
    \toprule
    Decomposition Unit & Representative Work & Prediction Approach \\
    \midrule
    GPU kernel 
      & NeuSight~\cite{neusight}\newline Habitat~\cite{habitat}
      & Estimate per-kernel execution time and sum \\
    \addlinespace
    Logical unit\newline(operator / layer)
      & Vidur~\cite{vidur}\newline MaverIQ~\cite{maveriq}
      & Extrapolate per-unit profiling results and sum \\
    \bottomrule
  \end{tabularx}
\end{table}


\section{Challenges of Latency Prediction in NPUs}
\label{sec:challenges}

Latency prediction is a critical capability for operating NPU-based serving systems, yet three challenges complicate it. These challenges preclude both the adaptation of existing NPU studies to the prediction task and the application of prior GPU prediction methodologies to NPUs.

\subsection{Undisclosed Microarchitecture of NPUs}
\label{sec:challenge_microarch}
Most prior NPU research has focused on simulation tools for design space exploration (DSE) of new NPU hardware~\cite{pytorch-sim, onnx-sim, scale-simv3, timeloop}. As a representative example, PyTorchSim~\cite{pytorch-sim} converts PyTorch-level model definitions to MLIR via Aten IR and performs cycle-level simulation on a systolic array-based virtual NPU, achieving a mean absolute error of 11.5\% when validated against the TPU v3 architecture. Similarly, ONNXim~\cite{onnx-sim} performs cycle-level simulation on ONNX graphs and enables DSE of multi-core NPUs at simulation speeds 384$\times$ faster than prior accelerator simulators.

\begin{table}[t]
  \centering
  \caption{\textbf{Microarchitectural disclosure at simulation grade across commercial NPUs.} 
  Each row is an input category required by prior NPU simulators~\cite{mnpusim, scale-simv3, pytorch-sim}.
  }
  \label{tab:npu_microarchitecture}
  \footnotesize
  \setlength{\tabcolsep}{4pt}
  \begin{tabular}{@{}lcccc@{}}
    \toprule
    & \multicolumn{3}{c}{Google TPU} & AWS \\
    \cmidrule(lr){2-4} \cmidrule(lr){5-5}
    Input Category 
      & v4~\cite{tpuv4-2023}
      & v5e~\cite{gcp-tpuv5e}
      & v6e~\cite{gcp-tpuv6e}
      & Inferentia2~\cite{aws-inferentia2} \\
    \midrule
    Systolic / MAC Array  & \cmark & \cmark & \cmark & \cmark \\
    Vector / Scalar Unit  & \xmark & \cmark & \xmark & \cmark \\
    On-chip Memory        & \xmark & \xmark & \xmark & \cmark \\
    Off-chip Memory (HBM) & \xmark & \xmark & \xmark & \xmark \\
    Interconnect          & \xmark & \xmark & \xmark & \xmark \\
    Clock Frequency       & \cmark & \xmark & \xmark & \xmark \\
    ISA / Prog. Model     & \xmark & \xmark & \xmark & \cmark \\
    \bottomrule
  \end{tabular}
\end{table}

However, such simulation-based approaches face structural limitations on NPUs, whose microarchitectural details are not fully disclosed. As shown in Table~\ref{tab:npu_microarchitecture}, the four NPUs commonly disclose only the dimensions and counts of their systolic arrays. None reveals the parameters essential for cycle-accurate simulation, such as cycle-level HBM timings or NoC routing policies.

To quantify the impact of this limitation, we conduct cycle-level simulation of Inferentia2 using PyTorchSim, with undisclosed fields filled by estimated values. While standalone MatMul is reproduced reasonably well with a 7.6\% error, element-wise and reduction operations exhibit errors as large as 232--323\%.
These errors stem from microarchitectural opacity~\cite{lew2019analyzing, pytorch-sim}. The authors of these simulators note that undisclosed microarchitectural parameters preclude exact cycle-count matching with the target hardware, and attribute the residual errors in their own validations to the same root cause.

\begin{figure*}[t]
    \centering
    \includegraphics[width=0.9\textwidth]{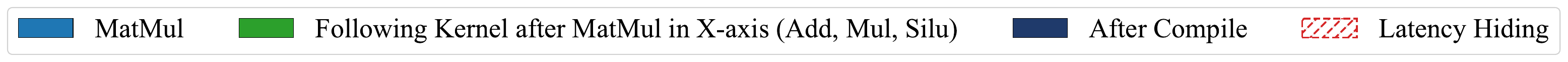}

    \vspace{0.5em}

    \subfloat[GPUs (T4, L4, A10G)\label{fig:hiding-gpu}]{%
        \includegraphics[width=0.49\textwidth]{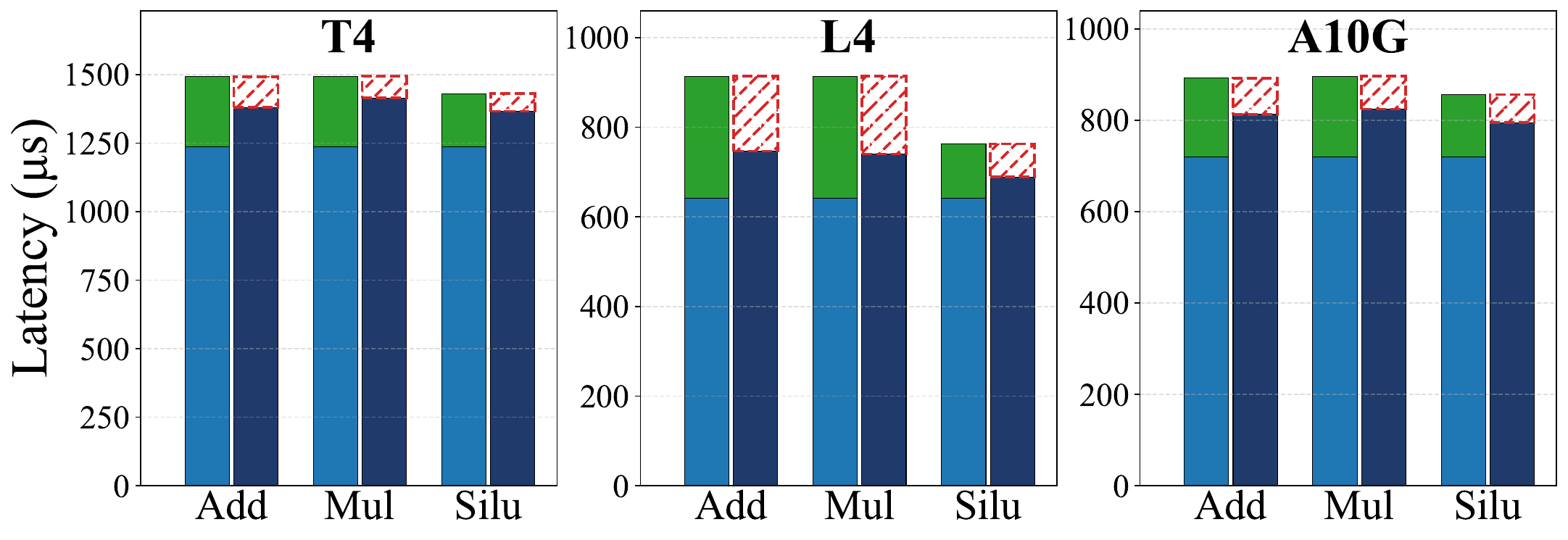}
    }
    \hfill
    \subfloat[NPUs (Inferentia2, TPU v5e, TPU v6e)\label{fig:hiding-npu}]{%
        \includegraphics[width=0.49\textwidth]{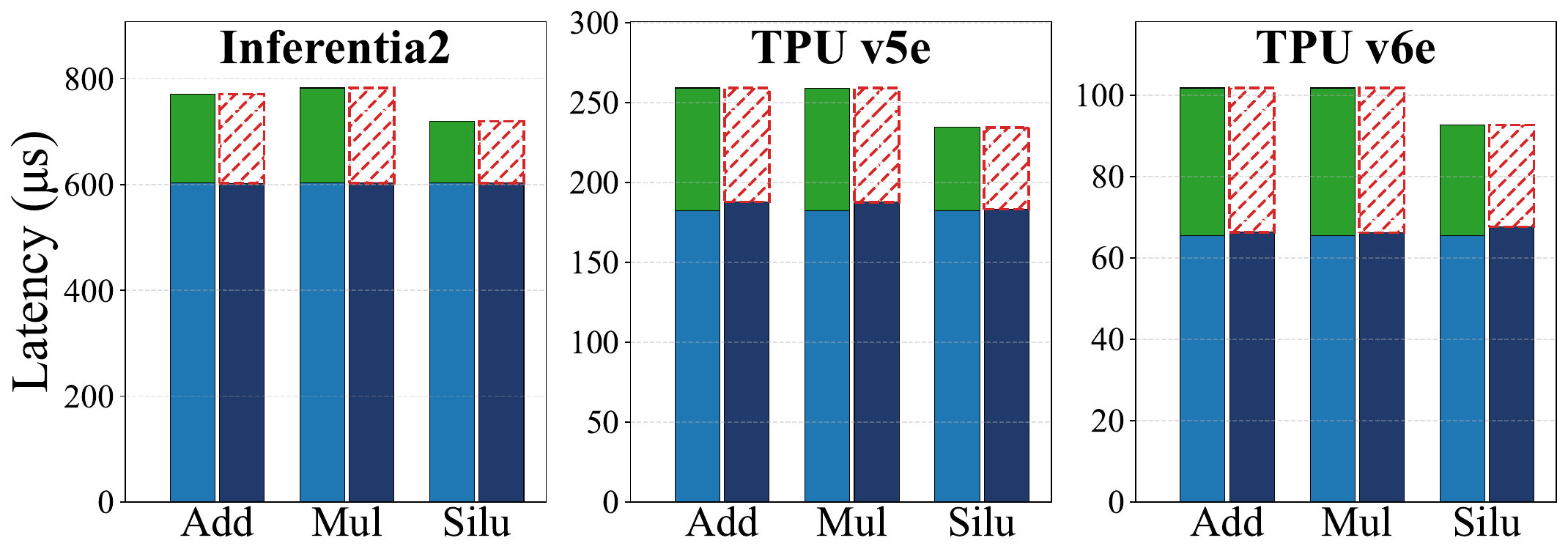}
    }

    \caption[Kernel fusion difference between NPUs and GPUs]{\textbf{Kernel fusion difference between NPUs and GPUs.} Blue bars show MatMul latency, green bars show the following kernel's latency (operation type on the x-axis). The stacked bars sum the individually measured latencies, while ``After Compile'' bars show the latency of the two operations executed back-to-back. The red hatched boxes indicate the latency reduction from compiler optimization.}
    \label{fig:hiding-experiment}
\end{figure*}

\begin{figure}[t]
    \centering
    \includegraphics[width=0.95\columnwidth]{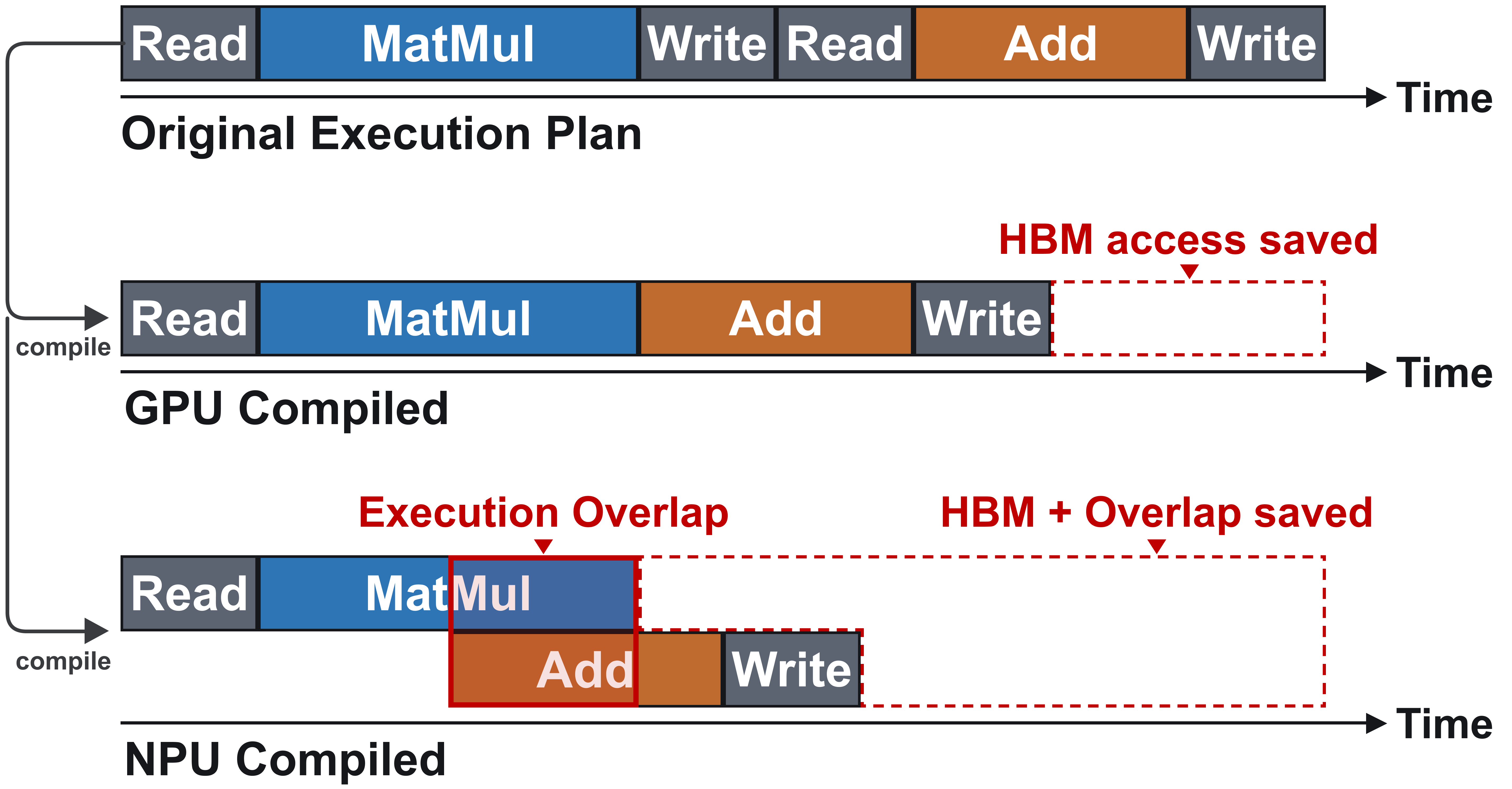}
    \caption[Compilation difference between GPUs and NPUs.]{\textbf{Compilation difference between GPUs and NPUs.}
    On GPUs, compiler optimization saves only the HBM access time between two kernels. On NPUs, the compiler can schedule the two operations to execute concurrently on separate engines.}
    \label{fig:kernel-fusion}
\end{figure}

\subsection{Unpredictable Compiler Optimizations}
\label{sec:challenge_compiler}

NPUs rely on vendor-provided compilers that lower model graphs to the underlying microarchitecture, applying graph optimizations such as kernel fusion, memory allocation, and execution scheduling. Compiler optimization is not unique to NPUs, but its scope on NPUs differs sharply from that on GPUs. On GPUs, automatic optimization rarely goes beyond fusing adjacent operations into a single kernel to reduce HBM round-trips~\cite{zhao2026neptune}, unless developers introduce hand-written fusion kernels such as FlashAttention~\cite{flash_attention}. On NPUs, by contrast, matrix, vector, and scalar operations run on distinct engines, so the compiler can schedule adjacent operations to execute concurrently. As Figure~\ref{fig:kernel-fusion} shows, such concurrent execution can hide one kernel's latency within another's~\cite{tpu-v1, TPU-v4i, tpuv4-2023}.

We measure this difference with a controlled experiment that chains a MatMul on the Tensor Engine with a subsequent operation (Add, Mul, or SiLU) on a separate engine through a data dependency. For each setup, we measure the standalone latencies of the two operations and the latency of their chained execution. As Figure~\ref{fig:hiding-experiment} shows, on NPUs the latency of the kernel following MatMul is nearly hidden and the total latency stays close to that of MatMul alone, whereas on GPUs the second kernel's latency is only partially reduced and no comparable hiding occurs. If even two operations produce such divergent behavior, predicting compiler optimizations across the hundreds of operations in an LLM inference graph is non-trivial.

The unpredictability of NPU compiler optimizations prevents the transfer of GPU-based prediction methodologies~\cite{neusight, habitat, vidur, maveriq} to NPUs. These GPU-based approaches commonly assume that individual kernels execute sequentially on the device, an assumption empirically supported on GPUs~\cite{daydream}. On NPUs, however, the compiler optimizations applied during per-unit profiling do not necessarily match those applied during E2E execution. This mismatch arises regardless of whether the decomposition unit is a kernel or a logical operator. Table~\ref{tab:e2e_comparison} reports the prediction errors of these methodologies when applied to NPUs.


\begin{table}[t] 
    \centering
    \caption[Prediction errors of GPU-based methodologies when applied to NPUs]{\textbf{Prediction errors of GPU-based methodologies when applied to NPUs.} Each column corresponds to a GPU-based latency prediction methodology, and each row corresponds to an input/output length configuration.}
    \label{tab:e2e_comparison}
    \small
    \setlength{\tabcolsep}{3pt}
    \begin{tabular}{lccc}
    \toprule
    Input/Output & \makecell{Kernel-sum\\ Error (\%)\cite{neusight,habitat}} & \makecell{Operator-sum\\ Error (\%)\cite{vidur}} & \makecell{Layer-sum\\ Error (\%)\cite{maveriq}} \\
        \midrule
    512/512    & $+80.9$  & $-63.58$ & $+22.02$ \\
    1024/1024  & $+493.3$ & $-60.64$ & $-43.00$ \\
    2048/2048  & $+446.0$ & $-57.37$ & $-11.35$ \\
    \bottomrule
    \end{tabular}
\end{table}

\subsection{Non-linear Latency Induced by Bucketing}
\label{sec:challenge_bucketing}

\begin{figure}[t]
\centering
\includegraphics[width=\columnwidth]{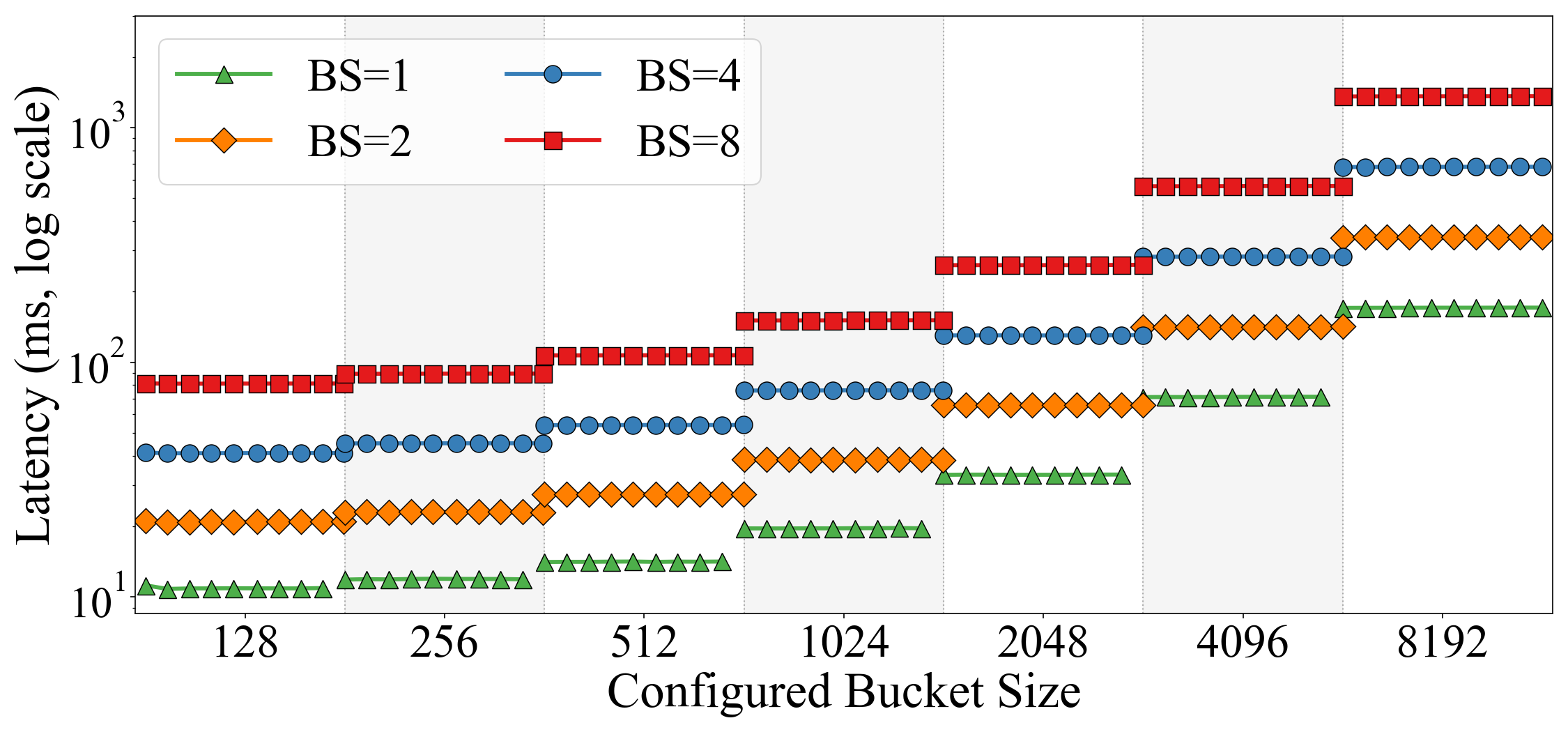}
\caption[Step latency induced by the bucketing effect on NPUs.]{\textbf{Step latency induced by the bucketing effect on NPUs.} TTFT measured per bucket on TPU v6e with Llama 3.1 8B deployed in a TP4 configuration, while varying the batch size across 1, 2, 4, 8. Each point corresponds to a different input length assigned to the bucket.}
\label{fig:bucketing_latency}
\end{figure}

NPUs support only static compilation, requiring a separate compilation for every input shape. However, LLM inference involves requests with varying input and output lengths, making per-length compilation impractical. To address this, NPU-based LLM serving systems adopt a \textit{bucketing} strategy: the compiler emits binaries only for a set of predefined lengths (e.g., 256, 512, 1024, 2048, ...), and incoming inputs are padded up to the nearest bucket size. Each bucket corresponds to a separately compiled binary, and inputs assigned to the same bucket are executed by the same binary. Figure~\ref{fig:bucketing_latency} plots the TTFT against input length, with each point measured at a different input length within the same bucket. Latency follows a step-function pattern, remaining constant within each bucket and jumping sharply at bucket boundaries. The overall LLM inference latency on NPUs is therefore governed by the sequence of buckets traversed during inference.

This bucketing structure carries two important implications for LLM inference latency prediction on NPUs. First, prediction approaches that assume a continuous latency function and interpolate between measurement points break down under this behavior. For example, MaverIQ~\cite{maveriq} interpolates LLM inference latency from a small number of measurement points, but this continuity assumption fails at bucket boundaries. As shown in Figure~\ref{fig:bucketing_latency}, a single-token increase across the 4096-to-8192 boundary multiplies TTFT by approximately 2.4× (e.g., from 71 ms to 171 ms at batch size 1), well outside the range that interpolation-based methods can absorb.

Second, profiling through actual execution at each measurement point becomes substantially more expensive on NPUs than on GPUs. Prior work~\cite{vidur, llmservingsim2.0, maveriq} reports profiling costs of several to tens of hours per model in GPU environments, and NPUs add a separate compilation step whenever the input shape changes. The official JAX documentation~\cite{jax_jit_doc} similarly notes that the cumulative compilation cost can dominate the actual profiling time as the number of measurement points grows. A practical prediction system for NPUs therefore must explicitly exploit the bucketing structure to keep the number of measurement points tractable.
\medskip

Consequently, the challenges identified in this section imply three requirements for LLM latency prediction methodology on NPU systems:
\begin{itemize}[label={}, leftmargin=0em]
\item \textbf{R1.}\quad Prediction without information on the NPU microarchitecture or compiler behavior.
`\item \textbf{R2.}\quad Reflection of bucketing-induced latency non-linearity.
\item \textbf{R3.}\quad Efficient profiling despite static compilation cost.
\end{itemize}

\section{LENS: Latency Estimator for NPU Systems}
\label{sec:lens_framework}
To satisfy the requirements for prediction methodology on NPUs, we build LENS on two key observations.
First, the E2E latency of a compiled binary is measurable even when the microarchitecture and compiler are treated as black boxes, making it a viable unit of prediction (R1).
Second, lengths within a bucket share the same compiled binary, so each bucket can be characterized from a few measurements without additional compilation (R2, R3).

Building on these observations, we propose LENS, a latency estimator that predicts the E2E latency of LLM inference on NPUs. LENS profiles each bucket with only a few measurements and composes the resulting per-bucket values to predict the latency of arbitrary input-output length combinations.

\subsection{Prediction Model}
\label{sec:lens_model}

LLM inference consists of a prefill phase and a decode phase. Since the latency of each phase is determined by a different bucket, LENS models the E2E latency as the sum of the prefill and decode latencies.

\paragraph{Prefill}
In the prefill phase, the entire input sequence is processed at once, and its latency corresponds to the TTFT. On NPUs, the TTFT is determined not by the actual input length but by the bucket to which the input is assigned. We define $TTFT_b$ as the prefill latency of bucket $b$.

Let $B$ denote the set of bucket sizes configured on the NPU. Given an input length $l^{\text{in}}$, we define the function that assigns it to a bucket as:
\begin{equation}
\texttt{bucket}(l^{\text{in}}) = \min \{ b \in B \mid l^{\text{in}} \le b \}
\end{equation}
The TTFT of a single request is then:
\begin{equation}
\label{eq:lens_ttft}
T_{\text{prefill}} = \text{TTFT}_{\texttt{bucket}(l^{\text{in}})}
\end{equation}

\paragraph{Decode}
In the decode phase, tokens are generated one step at a time, and the per-step latency corresponds to the TBT (Time Between Tokens). While the prefill bucket is determined by the input length $l^{\text{in}}$, the decode bucket at each step is determined by the size of the KV cache at that step. Let $c_t$ denote the KV cache length at generation step $t$. Since $c_t$ equals the input length plus the number of tokens generated so far, it can be expressed as:
\begin{equation}
\label{eq:lens_kv}
c_t = l^{\text{in}} + t + 1, \quad 0 \le t < l^{\text{out}}
\end{equation}

We define $\text{TBT}_b$ as the per-step decode latency of bucket $b$. The bucket used at step $t$ is $\texttt{bucket}(c_t)$, and the latency of that step is $\text{TBT}_{\texttt{bucket}(c_t)}$. If the KV cache stays within a single bucket $b$ throughout decoding, the decode latency for $l^{\text{out}}$ tokens is:
\begin{equation}
\label{eq:lens_decode_simple}
T_{\text{decode}} = l^{\text{out}} \cdot \text{TBT}_b
\end{equation}

In practice, $c_t$ grows monotonically as decode proceeds, so $c_t$ eventually exceeds the upper bound of the current bucket and the system switches to the binary of the next bucket. The decode process is therefore divided into multiple segments, each with a different $\text{TBT}_b$. Let $K$ be the total number of segments, $b_j$ the bucket used in segment $j$, and $l_j^{\text{out}}$ the number of tokens generated within that bucket. The decode latency is then:
\begin{equation}
\label{eq:lens_decode}
T_{\text{decode}} = l_1^{\text{out}} \cdot \text{TBT}_{b_1} + l_2^{\text{out}} \cdot \text{TBT}_{b_2} + \cdots = \sum_{j=1}^{K} l_j^{\text{out}} \cdot \text{TBT}_{b_j}
\end{equation}

Combining prefill and decode, the LENS E2E latency prediction for a single request is:
\begin{equation}
\label{eq:lens}
T_{\text{E2E}} = T_{\text{prefill}} + T_{\text{decode}} = \text{TTFT}_{\texttt{bucket}(l^{\text{in}})} + \sum_{j=1}^{K} l_j^{\text{out}} \cdot \text{TBT}_{b_j}
\end{equation}

\subsection{Extension to Batched Inference}
\label{sec:lens_batch}
In practice, LLM serving systems batch multiple requests together to improve throughput. The single-request formulation introduced above therefore needs to extend to the batched setting.

\paragraph{Batched Prefill}
NPU serving frameworks process each request in a batch independently in its own bucket. Therefore, the batch prefill latency for a batch of $N$ requests $\mathcal{R} = \{r_1, r_2, \ldots, r_N\}$ is the sum of the per-request prefill costs:
\begin{equation}
\label{eq:lens_prefill}
T_{\text{prefill}}(\mathcal{R}) = \sum_{i=1}^{N} \text{TTFT}_{\texttt{bucket}(l_i^{\text{in}})}
\end{equation}

\paragraph{Batched Decode}
Unlike prefill, where each request uses its own bucket, decode executes the entire batch with a single fixed shape. The TBT at each step is therefore determined by the bucket containing $c_t^{\max}$, the maximum KV position referenced by the batch:
\begin{equation}
\label{eq:lens_kv_batch}
c_t^{\max} =  \max_{i \in \mathcal{R}} c_t^{(i)}
\end{equation}

Furthermore, requests with shorter outputs are padded and executed alongside the others until the longest output completes, so the range of $t$ is determined by the longest output length in the batch:
\begin{equation}
0 \le t < \max_{i \in \mathcal{R}} l_i^{\text{out}}
\end{equation}

In the single-request case, we denoted the number of tokens generated within bucket $b_j$ as $l_j^{\text{out}}$. In the batched setting, however, the timing of bucket transitions is determined not by individual output lengths but by the trajectory of $c_t^{\max}$ across the entire batch. Let $n_j$ denote the number of steps that $c_t^{\max}$ stays within bucket $b_j$. The batched decode latency is then:
\begin{equation}
\label{eq:lens_decode_batch}
T_{\text{decode}}(\mathcal{R}) = \sum_{j=1}^{K} n_j \cdot \text{TBT}_{b_j}
\end{equation}

Combining prefill and decode, the LENS batched E2E latency prediction is:
\begin{equation}
\label{eq:lens_batch}
\begin{aligned}
T_{\text{E2E}}(\mathcal{R}) &= T_{\text{prefill}}(\mathcal{R}) + T_{\text{decode}}(\mathcal{R}) \\
&= \sum_{i=1}^{N} \text{TTFT}_{\texttt{bucket}(l_i^{\text{in}})} + \sum_{j=1}^{K} n_j \cdot \text{TBT}_{b_j}
\end{aligned}
\end{equation}

\paragraph{Extension to Optimized Attention Kernels}
Recent NPU serving frameworks introduce optimized attention kernels that allow each request to perform attention computation up to its actual KV length~\cite{rpa}, mitigating padding inefficiency. The overall computation still follows bucketing-based static compilation. We approximate decode latency by assigning each request to the bucket matching its own KV length.

While Eq.~\ref{eq:lens_decode_batch} assigns a single bucket's TBT to the entire batch, the batch still shares a single forward step under these kernels. The per-request cost is therefore the bucket's TBT divided by the batch size $N$.
\begin{equation}
\label{eq:lens_tbt_norm}
\overline{\text{TBT}}_b = \frac{\text{TBT}_b}{N}
\end{equation}

Each request then traverses buckets according to its own KV length, identical to the single-request model (Eq.~\ref{eq:lens_decode}). Applying Eq.~\ref{eq:lens_decode} to each request with the normalized TBT, the batched decode latency becomes:
\begin{equation}
\label{eq:lens_decode_rpa}
T_{\text{decode}}(\mathcal{R}) = \sum_{i=1}^{N} \sum_{j=1}^{K_i} l_j^{\text{out},(i)} \cdot \overline{\text{TBT}}_{b_j^{(i)}}
\end{equation}
where $K_i$ is the total number of buckets traversed by request $i$, $b_j^{(i)}$ is the $j$-th bucket of request $i$, and $l_j^{\text{out},(i)}$ is the number of tokens generated within that bucket.

\begin{table*}[t]
\caption{Dataset statistics before and after 8K-token filtering.}
\label{tab:dataset}
\small
\resizebox{\textwidth}{!}{%
\begin{tabular}{llr rrr rrr r}
\toprule
& & & \multicolumn{3}{c}{\# prefill tokens} & \multicolumn{3}{c}{\# decode tokens} & P:D \\
\cmidrule(lr){4-6} \cmidrule(lr){7-9}
Dataset & Content & \# queries & mean & median & p90 & mean & median & p90 & ratio \\
\midrule
ShareGPT~\cite{sharegpt} & Real-world ChatGPT conversations
  & 93K & 167 & 23 & 403 & 327 & 296 & 666 & 1\,:\,2 \\
\;(ShareGPT) & ShareGPT with max 8k total tokens
  & 93K & 167 & 23 & 403 & 327 & 296 & 666 & 1\,:\,2 \\
\addlinespace
CNN/DailyMail~\cite{cnndm} & News article summarization
  & 312K & 865 & 788 & 1{,}458 & 64 & 60 & 95 & 14\,:\,1 \\
\;(CNN) & CNN/DailyMail with input length $>$ 1{,}024 tokens
  & 92K & 1{,}393 & 1{,}305 & 1{,}872 & 73 & 67 & 113 & 19\,:\,1 \\
\addlinespace
ArXiv Summarization~\cite{arxiv-dataset} & Summarization of arxiv papers
  & 216K & 8{,}605 & 6{,}879 & 16{,}078 & 355 & 202 & 411 & 24\,:\,1 \\
\;(ArXiv) & ArXiv-Summarization with max 8k total tokens
  & 121K & 4{,}822 & 4{,}894 & 7{,}229 & 207 & 175 & 326 & 23\,:\,1 \\
\addlinespace
Writing Prompts~\cite{writingprompts} & Creative story generation from short prompts
  & 303K & 30 & 28 & 52 & 701 & 577 & 1{,}368 & 1\,:\,23 \\
\;(Writing Prompts) & Writing Prompts with max 8k total tokens
  & 303K & 30 & 28 & 52 & 701 & 577 & 1{,}368 & 1\,:\,23 \\
\bottomrule
\end{tabular}%
}
\end{table*}


\subsection{Profiling Point}
\label{sec:lens_profiling}

In the formulation of Eq.~\ref{eq:lens_batch}, the only measured quantities required by LENS are $\text{TTFT}_b$ and $\text{TBT}_b$ for each bucket $b$. Since the same binary executes for all input lengths within a bucket, both values can be obtained from only two E2E execution-time measurements per bucket. Specifically, for input length $l^{\text{in}}=b$, we measure the E2E latency with two different output lengths $L_1$ and $L_2$, yielding:
\begin{equation}
T_{\text{E2E}}^{(1)} = \text{TTFT}_b + L_1 \cdot \text{TBT}_b, \quad T_{\text{E2E}}^{(2)} = \text{TTFT}_b + L_2 \cdot \text{TBT}_b
\end{equation}
Solving this system yields:
\begin{equation}
\text{TBT}_b = \frac{T_{\text{E2E}}^{(2)} - T_{\text{E2E}}^{(1)}}{L_2 - L_1}, \quad \text{TTFT}_b = T_{\text{E2E}}^{(1)} - L_1 \cdot \text{TBT}_b
\end{equation}
As a result, two measurements per bucket suffice to predict the latency for all input lengths within that bucket. 



\section{Evaluation}
\label{sec:evaluation}

In this section, we validate that LENS achieves high prediction accuracy for arbitrary input-output length combinations from only a small number of profiling measurements. The evaluation covers four axes of variation. We test three models ranging from 1B to 14B parameters across four NPUs with different SDKs and architectures, sweep batch sizes from 1 to 32, and use four real-world datasets with diverse input-output length distributions.

\subsection{Experimental Setup}
\label{sec:eval_setup}

\paragraph{NPU}
LENS's prediction approach is general and does not depend on any specific NPU or compiler stack. On AWS, we use Inferentia2 with the Neuron SDK, and on GCP, we use TPU v4, TPU v5e, and TPU v6e with JAX/XLA. These NPUs span a wide range of peak compute (190--918 BF16 TFLOPS) and HBM bandwidth (0.8--1.6 TB/s), while sharing the common characteristics of heterogeneous engine design and static compilation. All experiments use a 4-chip configuration, which corresponds to TP4 on TPUs, and to TP8 on Inferentia2 since it has 2 cores per chip.

\paragraph{Models}
To ensure diversity in computational structure, we use three models from different families: Llama-3.2-1B, Mistral-7B-v0.3, and Qwen3-14B, covering parameter scales from 1B to 14B. We vary the batch size from 1 to 32, subject to each hardware's memory capacity.

\paragraph{Datasets}
To evaluate on diverse input-output workloads, we use four datasets. ShareGPT~\cite{sharegpt} consists of real-world ChatGPT conversations with short inputs but a long-tailed distribution. CNN/DailyMail~\cite{cnndm} is a news summarization dataset, from which we extract only samples with input length above 1{,}024 tokens to cover the mid-length range. ArXiv Summarization~\cite{arxiv-dataset} consists of academic papers with very long inputs (mean of 4{,}808 tokens). Writing Prompts~\cite{writingprompts} is a creative writing workload that generates long outputs from short inputs (mean output of 697 tokens), covering the long-output regime. The input-output length distributions of each dataset are summarized in Table~\ref{tab:dataset}. Samples exceeding 8,192 tokens in maximum sequence length are excluded to fit within each NPU's HBM capacity.

\paragraph{Measurement Setup}
We measure latency using the native serving framework of each NPU platform: NxDI on AWS Inferentia2 (Neuron SDK), and MaxText on GCP TPUs (JAX/XLA). Both frameworks execute LLM inference under the compiler's full optimization pipeline, matching the bucket-level E2E latency that LENS predicts. We use a fixed bucket set $B = \{128, 256, 512, 1024, 2048, 4096, 8192\}$ across all experiments.


\begin{figure*}[!t]
\centering
\includegraphics[width=\textwidth]{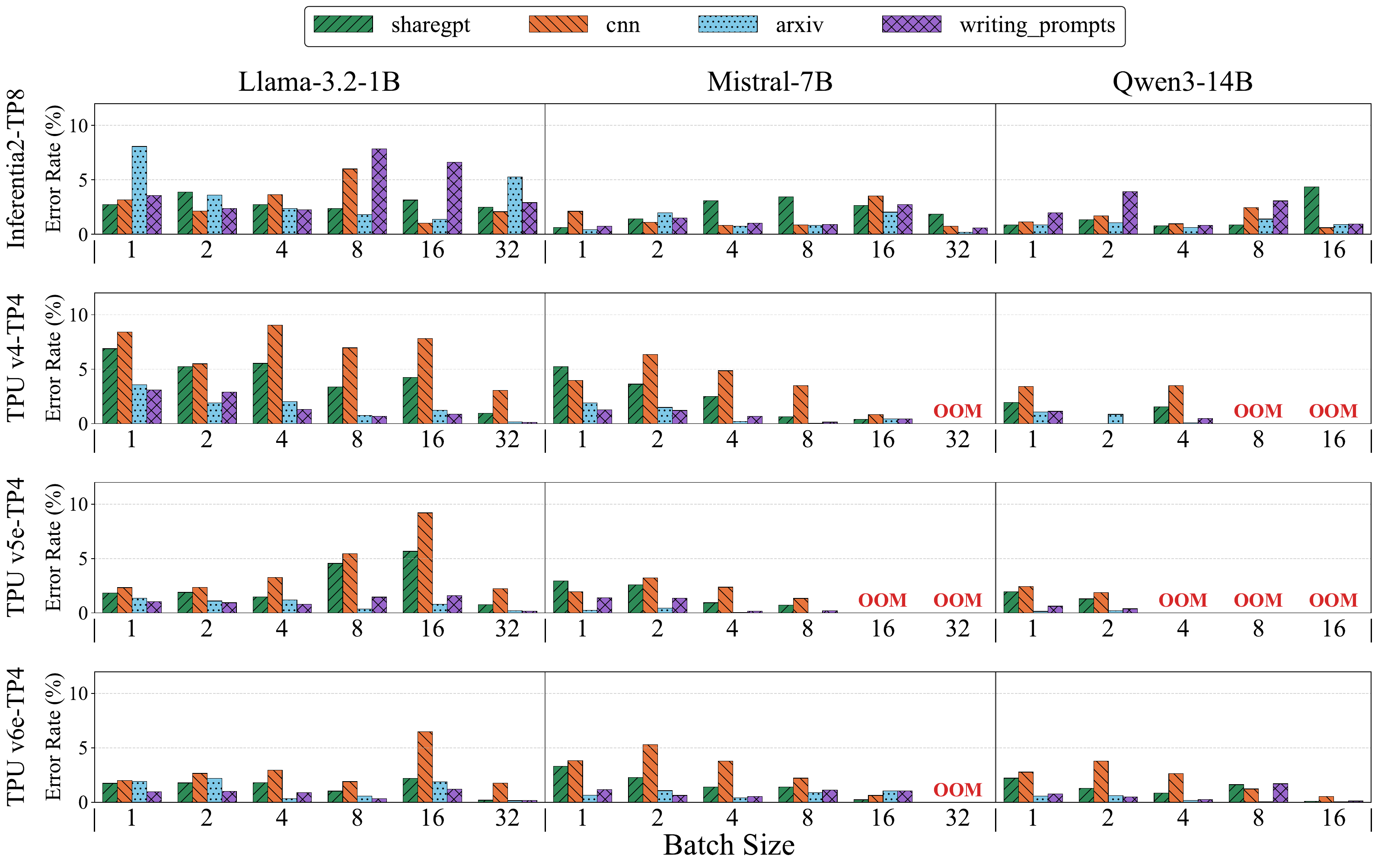}
\caption[Accuracy evaluation of LENS on four NPUs]{\textbf{Accuracy evaluation of LENS on four NPUs.} Each bar reports the prediction error against measured latency.}
\label{fig:evaluation}
\end{figure*}


\subsection{Prediction Accuracy in Diverse Settings}
\label{sec:eval_results}

Figure~\ref{fig:evaluation} reports LENS's E2E latency prediction errors across four NPU configurations (all with 4 chips), three models, and four datasets. The main evaluation covers 248 cases spanning four NPUs, three models, four datasets, and multiple batch sizes. LENS achieves a mean error of 2.15\% (median 1.69\%), with 89.9\% of cases falling under 5\% error.

\paragraph{Consistency across NPUs}
LENS maintains low error rates across all four NPUs, despite their different compiler stacks and hardware architectures. The mean errors are 2.20\% (max 8.07\%) on Inferentia2-TP8, 2.65\% (max 9.05\%) on TPU v4-TP4, 2.35\% (max 11.47\%) on TPU v5e-TP4, and 1.46\% (max 6.52\%) on TPU v6e-TP4. These results indicate that bucket-based modeling remains effective across both compiler stacks (Neuron SDK and JAX/XLA) and hardware architectures with different systolic array dimensions and memory hierarchies.

\paragraph{Consistency across model scales}
LENS maintains stable accuracy from 1B (Llama-3.2-1B) to 14B (Qwen3-14B). The mean errors are 2.70\% for Llama-3.2-1B, 1.61\% for Mistral-7B, and 1.52\% for Qwen3-14B, with errors decreasing as model size and batch size grow.

\paragraph{Consistency across workload characteristics}
LENS maintains accurate predictions across datasets with diverse input-output length distributions. The mean errors are 1.21\% on ArXiv (prefill-dominant) and 1.41\% on Writing Prompts (decode-dominant), indicating that bucket-based modeling captures both extremes of the prefill-decode spectrum. ShareGPT and CNN/DailyMail show slightly higher mean errors of 2.49\% and 3.49\%, respectively. The shorter sequences in these datasets yield smaller absolute latencies, where measurement noise forms a larger fraction of the prediction error. All four datasets remain within 5\% on average, confirming that dataset characteristics have little impact on prediction accuracy.


\subsection{Comparison with Baselines}
\label{sec:eval_baseline}
Direct baselines for NPU latency prediction are scarce, so we select two methodologically related works.
LLM Serving Sim 2.0~\cite{llmservingsim2.0} is a unified LLM serving simulator validated on NPUs. It sums per-operator latencies and applies an E2E correction step to mitigate the black box nature of compiler optimization, a strategy methodologically aligned with LENS.

MaverIQ~\cite{maveriq} is a layer-level latency predictor for GPU-based LLM serving. It models latency as a linear function of token count, with a single TTFT constant and a single TBT slope across the entire dataset. LENS adopts a similar linear assumption but applies it per bucket, making MaverIQ a methodological baseline despite its GPU origin and global linearity.

Since LLM Serving Sim 2.0 is built on the vLLM-TPU codebase, we use vLLM-TPU as the serving framework for all methods.

\paragraph{Baseline-Results}
LLM Serving Sim 2.0 and MaverIQ both achieve reasonable accuracy in certain regions but reveal large errors where their assumptions break.

LLM Serving Sim 2.0 errors amplify as batch size grows and prefill operations accumulate. On ArXiv under BS=32, the error exceeds 200\%. Despite the E2E latency correction step, per-operator summation fails to capture graph-level compiler optimizations.

MaverIQ assumes no bucketing. Fitting a single TTFT constant and TBT slope to an entire dataset, it incurs sharply growing errors on workloads where bucket transitions occur frequently during inference. On CNN, errors within the same dataset split between regions exceeding 80\% and regions in the 40\% range depending on batch size, and on ShareGPT, errors of 50--70\% persist across all batch sizes. MaverIQ averages 33\% Mean Absolute Percentage Error (MAPE) across all configurations.

LENS addresses both limitations. It uses E2E measurement as the unit of prediction, which captures compiler optimizations that per-operator summation misses, and it maintains a separate model per bucket, which captures the discontinuity at bucket transitions that a global linear fit cannot express. On the same configurations, LENS achieves an average MAPE of 5.3\%.


\begin{figure*}[!t]
\centering
\includegraphics[width=\textwidth]{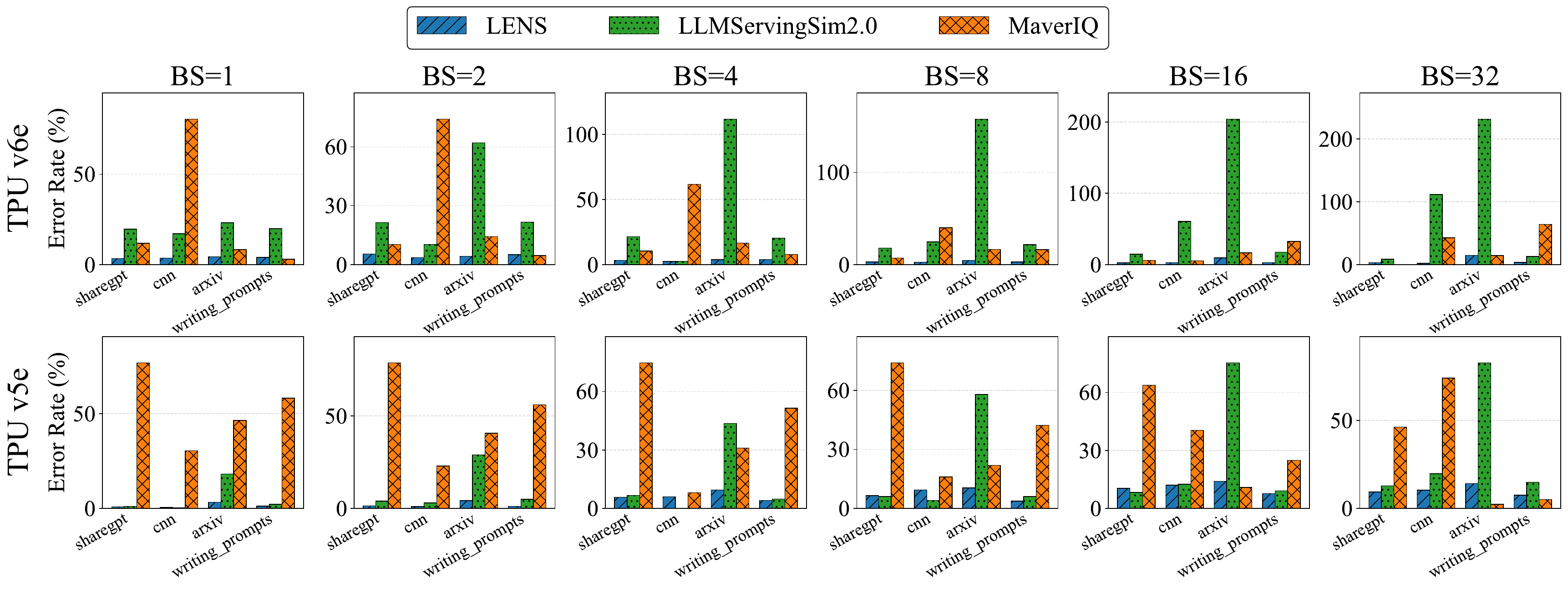}
\caption{Comparison with Baseline}
\label{fig:baseline}
\end{figure*}

\paragraph{Profiling Cost Comparison}
LLM Serving Sim 2.0 demands the largest profiling budget. Per-operator profiling requires 5{,}000 measurement points, and each point requires a separate compilation. Total profiling time, including compilation, reaches 7 hours.

MaverIQ uses a single decoder layer as the profiling unit, which shortens execution time compared to E2E measurement. However, each dataset requires a separate compilation graph, leading to 4 compilations per batch and 41 minutes of compilation needed across the experiment. This compilation cost dominates MaverIQ's total profiling time of 45 minutes, leaving the layer-level profiling unit poorly matched to the actual cost structure.

LENS uses 14 profiling points per batch, but only 1 compilation per batch. The entire configuration requires 12 compilations in total. Compilation takes 10 minutes, and the total profiling time including execution is 18 minutes.

Reducing the profiling unit is therefore not sufficient on NPUs. Profiling cost is dominated by compilation, so the design must minimize the number of compilations. LENS uses more profiling points than MaverIQ but requires only 12 compilations against MaverIQ's 48, matching the cost structure of NPU profiling.



\subsection{Case Study: Where Intuitions Fail on NPUs}
\label{sec:case_study}

GPU intuitions about LLM serving do not carry over to NPUs. 
The compiler reshapes execution per configuration in ways that cannot be anticipated. NPU serving system operators therefore cannot reason about an optimum without measuring it. LENS serves as a key tool: it needs only a few profiling points, an advantage amplified on NPUs where compile time dominates profiling cost. The two case studies below illustrate this role.


\begin{figure}[t]
    \centering
    \includegraphics[width=0.95\linewidth]{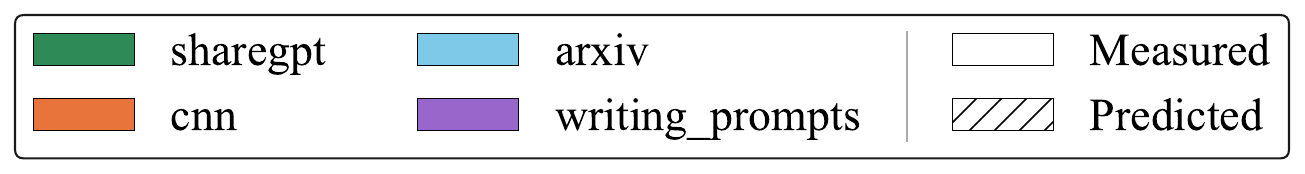}\\
    
    \subfloat[Mistral 7B]{%
        \includegraphics[width=0.48\linewidth]{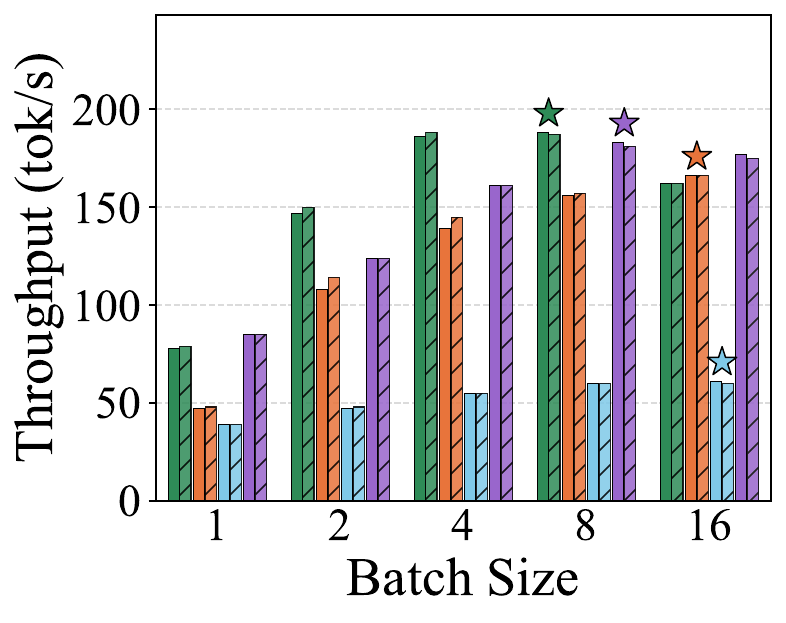}%
        \label{fig:sweat_spot_mistral_7b}}
    \hfil
    \subfloat[Qwen 3 14B]{%
        \includegraphics[width=0.48\linewidth]{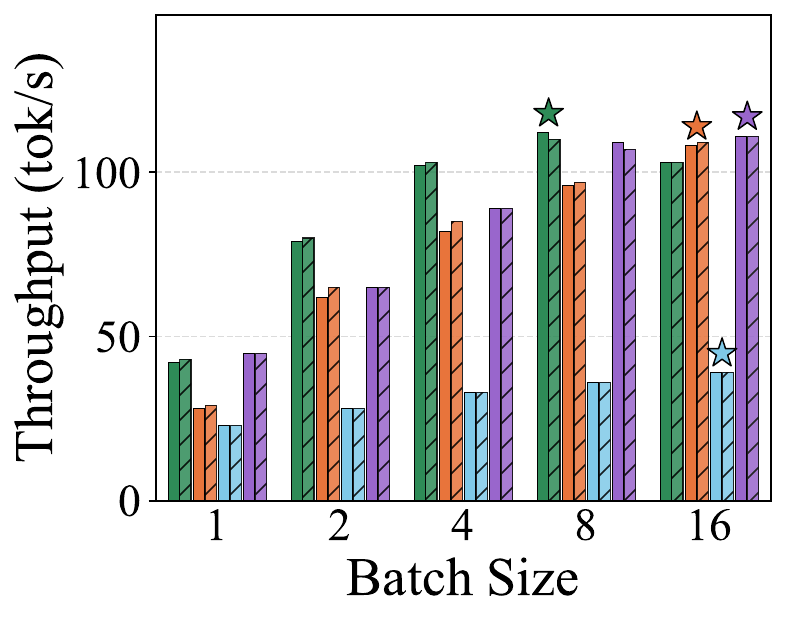}%
        \label{fig:sweat_spot_qwen_14b}}
    
    \caption{Measured and predicted throughput on TPU v6e across batch sizes for each dataset. Stars mark the batch size that achieves the highest throughput for each dataset.}
    \label{fig:plot_both}
\end{figure}

\subsubsection{Case Study 1: Larger Batches Do Not Always Maximize Throughput on NPUs}
\label{sec:case_tradeoff}

Larger batches on GPUs raise throughput at the cost of per-request latency~\cite{orca, vllm, mind-the-memory-gap}. Throughput climbs with batch size and then plateaus once the GPU saturates.

NPUs break this pattern. The NPU compiler is shape-sensitive, and the execution path it builds for a given input shape can be either highly optimized or far less efficient, depending on the shape itself~\cite{rpa}. The GPU intuition that larger batches always raise throughput therefore no longer holds on NPUs. Each batch size produces a distinct input shape, and the compiler may optimize one shape but not a neighboring one. A larger batch can therefore land on a slower path and deliver lower throughput than a smaller one.

Figure~\ref{fig:plot_both} illustrates this across two models. On TPU v6e with Mistral 7B, the throughput-maximizing batch size shifts with the dataset (Figure~\ref{fig:sweat_spot_mistral_7b}): CNN and ArXiv peak at BS 16, while ShareGPT and Writing Prompts peak at BS 8. The same pattern appears on Qwen3 14B (Figure~\ref{fig:sweat_spot_qwen_14b}): most datasets peak at BS 16, but ShareGPT peaks at BS 8.

Each (LLM, NPU, dataset) combination thus requires its own search for the throughput-maximizing batch size. LENS makes this search practical, predicting these peaks accurately and locating them quickly, as Figure~\ref{fig:plot_both} confirms.

\begin{figure}[t]
    \centering  
    \includegraphics[width=0.98\columnwidth]{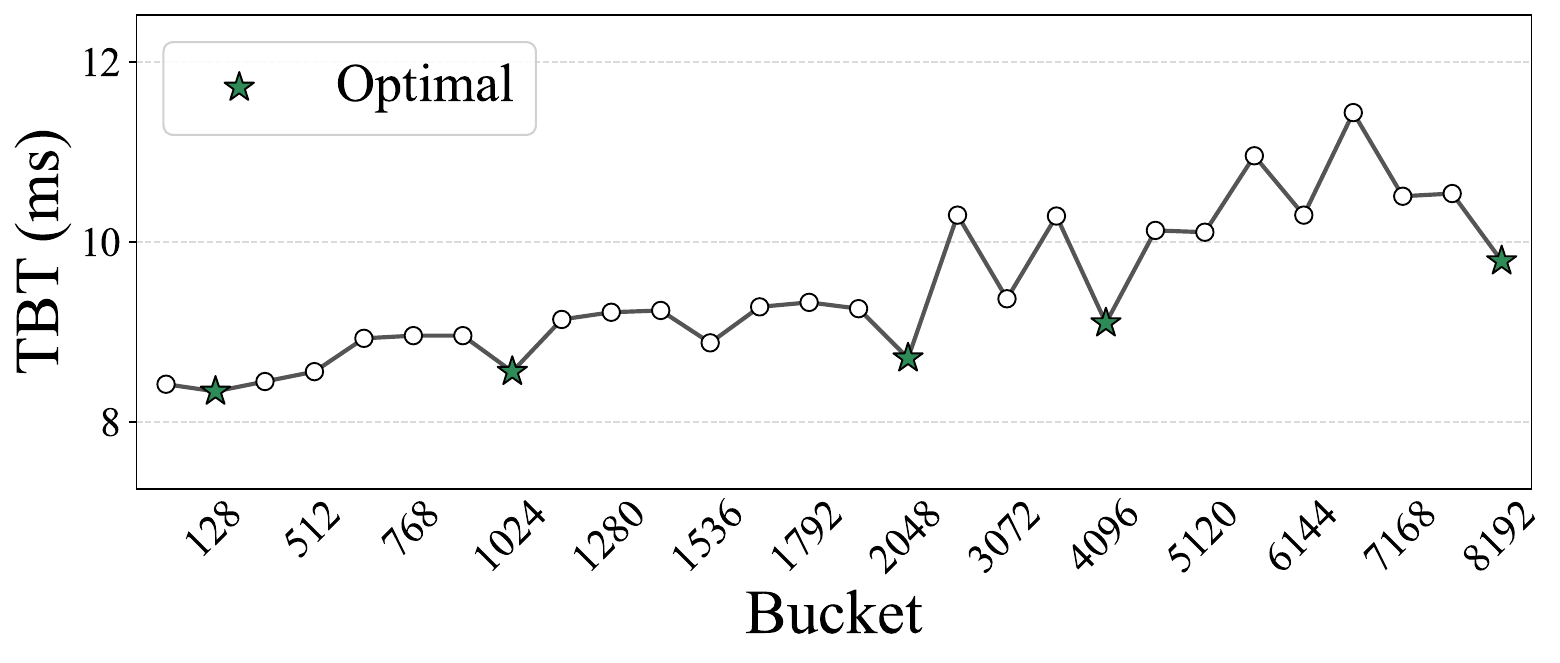}
    \caption{Measured TBT across buckets (Inferentia2, TP=2, Llama-3.2 1B, BS=1). Stars mark the buckets that lie on a compiler-optimized execution path.}
    
    \label{fig:case_study2}
\end{figure}

\subsubsection{Case Study 2: Smaller Buckets Do Not Always Run Faster on NPUs}
\label{sec:case_bucket}

Bucket selection is a key decision in NPU-based LLM serving, on par with tensor parallelism and batch size. Each compiled bucket occupies its own HBM buffer, so memory bounds the total bucket count. Operators must therefore select a small set tailored to the workload.

The natural intuition is that smaller buckets should run faster: a smaller bucket pads less, so each request does less work. For ArXiv dataset, this suggests subdividing the 4096--8192 range at 512-token steps to push most requests into smaller, lighter buckets, rather than relying on the default \{4096, 8192\}. In practice, the finer-grained set increases E2E latency by 7\% over the default.

The compiler optimizes only certain bucket shapes~\cite{rpa}, and an optimized larger bucket can be faster than an unoptimized smaller one (Figure~\ref{fig:case_study2}). Bucket latency on NPUs is therefore not monotonic in bucket size: it depends on which shapes the compiler chose to optimized path. The optimal bucket set must account for this compiler behavior as well as the input/output length distribution of the workload, leaving operators with an empirical search across the full bucket space.

Each new bucket still requires its own profiling. LENS reduces this from a full deployment per bucket combination to two E2E measurements per bucket, turning the search from an exhaustive campaign into a quick exploration that surfaces the buckets the compiler actually optimizes.

\paragraph{Summary}
Both case studies expose a common pattern: GPU-derived intuitions fail to predict NPU behavior, and the optimum can only be confirmed by direct measurement. The compiler and architecture differences that drive this also make analytical reasoning unreliable. LENS fills this gap as a practical search tool, letting operators explore the configuration space and identify the choices that yield the desired performance.

\section{Future Work}
\label{sec:discussion}

While LENS achieves efficient profiling with only two measurements per bucket, its predictions remain confined to the NPU on which profiling was conducted. NeuSight~\cite{neusight} and Habitat~\cite{habitat} address the analogous limitation in the GPU domain by decomposing inference into kernel-level units and applying analytical scaling factors derived from hardware characteristics such as compute throughput and memory bandwidth. LENS, however, operates on bucket-level E2E latency without kernel decomposition, rendering kernel-level scaling factors directly inapplicable. Whether bucket-level E2E latency itself exhibits a consistent scaling relationship across NPUs, and what set of hardware features most accurately predicts such a relationship, remains an open question. If such a relationship can be established, LENS could be extended to support prediction across heterogeneous NPU platforms without requiring profiling on each. We leave the empirical characterization of cross-NPU scaling behavior to future work.

LENS currently predicts latency under an offline serving setting. Production serving environments introduce system-level factors that affect E2E latency, such as queuing delay, inter-stage communication overhead in pipeline parallelism, and request routing across heterogeneous accelerators. Extending LENS to system-level performance prediction that accounts for these factors is a direction for future work.

\section{Related Work}
\label{sec:related}

\paragraph{DNN Performance Estimation}

A substantial body of prior work has investigated DNN performance prediction~\cite{habitat, neusight, vidur, life-framework, maveriq}. Habitat~\cite{habitat} predicts kernel execution time on a target GPU by applying an analytical wave scaling factor derived from the roofline model, and integrates an ML-based approach to capture non-linearity in GPU computation. Vidur~\cite{vidur} profiles the operators arising in LLM inference by partitioning them into three categories (i.e., token-level operators, sequence-level operators, communication operators) and predicts inference latency through random-forest-based interpolation. MaverIQ~\cite{maveriq} minimizes profiling cost via a proxy model that loads only the unique layers of an LLM, and predicts inference latency across diverse configurations through an analytical communication model.

However, these studies are predicated on GPU architectures, and naive extensions that fail to account for the architectural characteristics of NPU systems (e.g., bucketing) yield substantial error rates. Furthermore, since each configuration on NPUs incurs a separate compilation, the granularity-adjustment approach of existing GPU methodologies cannot effectively reduce profiling overhead. By accounting for all of these characteristics, LENS attains low error rates with only two profiling measurements per bucket.

\paragraph{NPU Prior Work}
Cycle-accurate simulators for NPU performance analysis have been developed in various forms, including multi-core NPU resource sharing (mNPUsim~\cite{mnpusim}), systolic array accelerators (SCALE-Sim v3~\cite{scale-simv3}), and PyTorch-integrated frameworks (PyTorchSim~\cite{pytorch-sim}).

However, all of these simulators share the common assumption that users must provide microarchitecture specifications as input, making them difficult to apply directly to commercial NPUs where microarchitectural details are not sufficiently disclosed.
In addition, they do not explicitly address LLM inference characteristics such as the distinction between prefill and decode phases, or bucketing.
LENS treats the microarchitecture and compiler as a black-box and predicts LLM inference latency using only bucket-level E2E profiling, complementing these simulation-based approaches.

\section{Conclusion}
\label{sec:conclusion}
This paper presents LENS, a tool that accurately predicts LLM inference latency on NPU systems. We characterize three challenges that preclude prior work from being applied to LLM inference latency prediction on commercial NPUs: the undisclosed microarchitecture, the unpredictability of compiler optimizations, and the bucketing-induced latency non-linearity. LENS addresses these challenges by measuring only two E2E execution times per bucket and composing them to predict E2E latency for arbitrary input-output combinations.

Across 248 cases on four commercial NPUs (Inferentia2, TPU v4, v5e, v6e), three models, and four datasets, LENS achieves a mean error of 2.15\% and a median of 1.69\%, with 89.9\% of cases within 5\% error. We further evaluate LENS against two methodologically related baselines, LLM Serving Sim 2.0 and MaverIQ, selected to validate our design choices in a controlled comparison. Both baselines exhibit substantial errors where their assumptions break (over 200\% and over 70\% respectively), while LENS maintains consistent accuracy with an average MAPE of 5.3\%. Furthermore, two case studies demonstrate the need for LENS as a search tool for NPU serving operators facing configuration problems that defy GPU-derived intuitions.

\bibliographystyle{IEEEtranS}
\bibliography{Reference}

\end{document}